\documentclass[proof]{WileyASNA-v1}
\usepackage{epsfig}
\usepackage{upgreek}
\usepackage{graphicx} 
\usepackage{subfig}
\usepackage{tikz,pgfplots}
\articletype{Original Paper}%

\received{<day> <Month>, <year>}
\revised{<day> <Month>, <year>}
\accepted{<day> <Month>, <year>}

\begin{document}

\title{ Fe~II emission in active galactic nuclei}

\author[1]{C. Martin Gaskell*}

\author[1,2]{Neha Thakur}

\author[1,3]{Betsy Tian}

\author[1,4]{Anjana Saravanan}

\authormark{GASKELL \textsc{et al.}}

\address[1]{\orgdiv{Astronomy \& Astrophysics}, \orgname{University of California Santa Cruz}, \orgaddress{\state{CA 95064}, \country{USA}}}

\address[2]{\orgdiv{Data Science Department}, \orgname{University of California Berkeley}, \orgaddress{\state{CA 94720}, \country{USA}}}

\address[3]{\orgdiv{Physics Department}, \orgname{Massachusetts Institute of Technology}, \orgaddress{\state{MA 02139-4307}, \country{USA}}}

\address[4]{\orgdiv{Department of Mechanical Engineering}, \orgname{University of California Berkeley}, \orgaddress{\state{CA 94720-1740}, \country{USA}}}

\corres{*Corresponding author: C.~M.~Gaskell
\email{mgaskell@ucsc.edu}}


\abstract[Abstract]{We review problems raised by Fe~II emission in active galactic nuclei (AGNs) and address the question of its relationship to other broad-line region (BLR) lines.  The self-shielding, stratified, BLR model of Gaskell, Klimek \& Nazarova (2007; GKN) predicts that Fe~II emission comes from twice the radius of H$\upbeta$, in agreement with widths of lines of Fe~II lines being only 70\% of the widths of H$\upbeta$. This disagrees with some reverberation mapping results which have suggested that Fe~II and H$\upbeta$ arise at similar radii.  The highest quality reverberation mapping, however, supports the predictions that Fe~II comes from twice the radius of H$\upbeta$.  We suggest that lower quality reverberation mapping of Fe~II is biased to give too small lags.  We conclude that, in agreement with the GKN model, the region emitting Fe~II is the outermost part of the BLR just inside the surrounding dust.   The model naturally gives the Doppler broadening require by models of Fe~II emission. Optical Fe~II emission implies typical reddenings of $E(B-V) \sim 0.20$.  This helps explain the ratio of UV to optical Fe~II emission.  Simulations show that the amplitude of Fe~II variability is consistent with being the same as for H$\upbeta$ variability.  The Fe~II/H$\upbeta$ ratio is a good proxy for the Eddington ratio.  The ratio might be driven in part by the strong soft X-ray excess because the X-rays destroy grains and release iron into the gas phase.  We propose that the correlation of Fe~II strength of radio and host galaxy properties is a result of AGN downsizing. }


\jnlcitation{\cname{%
\author{Gaskell, C.~M.}, 
\author{Thakur, N.}, 
\author{Tian, B.}, and
\author{Saravanan, A}} (\cyear{2021}), 
\ctitle{Fe~II emission in active galactic nuclei}, \cjournal{Astron. Nachr.}, \cvol{2021;00:1--6}.}


\maketitle


\section{Introduction}\label{sec1}

``Understanding the Fe~II emission from Active Galactic Nuclei has been a grand challenge for many decades,'' write \citet{Sakar+21}.  This might seem to be a surprising statement since iron is one of the most abundant heavy elements in universe.  Of the elements heavier that oxygen, only neon has a significantly higher solar abundance.\footnote{The abundances of magnesium and silicon are similar to iron.}  Iron absorption lines dominate the spectrum of the sun, so much so that, prior to \citet{Payne1925}, it was believed that the sun was composed primarily of iron. Given its high cosmic abundance, it should be no surprise to find iron lines in the spectra of active galactic nuclei (AGNs).  However, trying to understand the origin and nature of Fe~II emission in AGNs has raised many questions.

Fe~II emission in AGNs was first identified by \citet{Wampler+Oke67} in 3C~273. The spectrum of Fe~II is very rich because Fe$^+$ has a large number of low-lying energy levels. In the optical, the most readily observable multiplets occur in broad blends around $\lambda$4570 and $\lambda$5300.  In 3C~273 as in most AGNs, the Doppler widths of the Fe~II lines are too broad for individual lines within multiplets to be resolved but \citet{Sargent68} discovered strong narrow Fe~II emission in I~Zw~1 where individual lines could be resolved. Sargent noted that, apart from the difference in line widths, the ratio of line strengths was similar in I~Zw~1 to 3C~273.  I~Zw~1 is the prototype of AGNs which have come to be called ``narrow-line Seyfert 1s'' or ``NLS1s'' \citep{Gaskell84,Osterbrock+Pogge85}.
We show a typical spectrum of a NLS1 in Figure 1. 

In this paper we review some of the problems associated with Fe~II emission in AGNs and argue that the region emitting Fe~II is part of the normal broad-line region (BLR) of AGNs. We will focus on Fe~II emission in the optical, since this is the most readily observed Fe~II emission, but it is also observed in the UV and the near IR.

\begin{figure}
\centering 
\includegraphics[width=0.48\textwidth,angle=0]{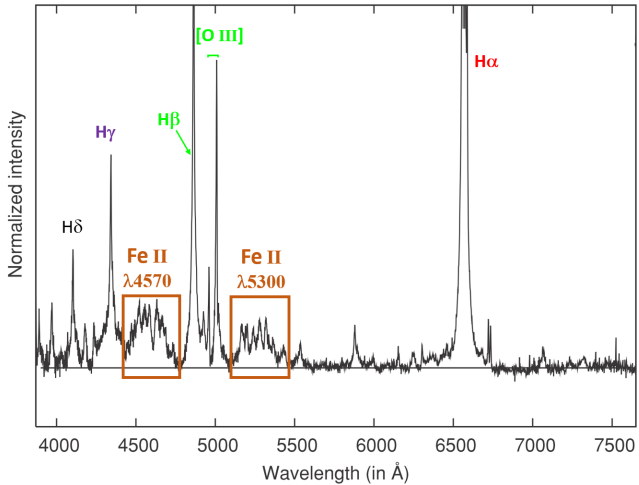}
\caption{A continuum-subtracted spectrum of the typical NLS1 showing Fe~II and other emission lines.  The square boxes indicate the $\lambda$4570 and $\lambda$5300 blends commonly taken as an indication of Fe~II strength.  (Figure adapted by permission from \citealt{Kovacevic+10})}.
\end{figure}

\section{Background}

\subsection{Fe~II in Galactic objects}

Optical Fe~II emission was first identified by \citet{Moore+Sanford1914} in the spectrum of $\upeta$ Carinae.  \citet{Wampler+Oke67} identified Fe~II in 3C\, 273 on the basis of the spectrum of Nova Herculis 1934 \citep{Stratton+Manning39}. Fe~II is also very prominent in the spectrum of RR Telescopii \citep{Thackeray53}.  The existence of strong Fe~II emission in these objects immediately tells us that {\it a black hole is not necessary for strong Fe~II emission.}

One important question that was {\it not} asked in the first half of the 20th century was, ``why don't planetary nebulae and H II regions show strong Fe~II emission?'' The strengths of Fe lines in planetary nebulae and H II regions are a couple of orders of magnitude weaker than one would expect \citep{Delgado-Inglada09}.   \citet{Howard+63} found that gas-phase calcium was depleted by two orders of magnitude and recognized that this was due to calcium going into grains.  \citet{deBoer+72} found a similarly large depletion iron.  It is thus now recognized that the weakness of Fe~II lines in planetary nebulae and H II region is because of depletion of refractory elements onto grains.  Conversely, the strength of Fe~II in objects like $\upeta$ Carinae and symbiotic novae implies that these objects have significant amounts of gas where iron is {\it not} strongly depleted onto grains.

A valuable thing about Galactic Fe~II emitters like RR Telescopii is that the Fe~II lines are far narrower than even in the most extreme NLS1.  Important insights into the microphysics of Fe~II emission have been gained because of this.

\subsection{Fe~II in 3C 273}

By the time of the identification of Fe~II in 3C~273 it was recognized that there was a dichotomy between two distinct emission-line systems in AGNs \citep{Dibai+Pronik67}: what we now call the ``narrow-line region'' (NLR) and the ``broad-line region'' (BLR). \citet{Dibai+Pronik67} estimated that the NLR had a density, $n_e \sim 10^3$ cm$^{-3}$ and a size of 100s of pc while the BLR had a density $n_e > 10^6$ cm$^{-3}$ and a size of $< 1$ pc.  The just-mentioned Galactic objects with strong Fe~II emission show forbidden Fe~II lines as well as permitted Fe~II lines.  \citet{Wampler+Oke67} failed to find the expected [Fe~II] lines in 3C\,273 and deduced from this that the density of the gas producing the Fe~II lines had $n_e > 10^7$ cm$^{-3}$.  This placed the Fe~II emission with the broad-line region gas, not in the NLR.

\citet{Wampler+Oke67} also considered the relative strengths of Mg II and Fe~II emission, lines that should arise under similar conditions.  They deduced that the Mg/Fe abundance ratio was consistent with being solar -- i.e., an unusually high Fe abundance was not needed.

\section{Seven remarkable things about optical Fe~II emission}

\subsection{Relative strength}

Fe~II emission in AGNs can be very strong.  Its total observed flux can readily exceed a quarter or more of the flux in all other lines.  Figure 2 shows the ratio of intensities of Fe~II $\uplambda$4570/H$\upbeta$.  It can readily be seen from inspection of Figures 1 and 2 that the flux of the $\uplambda$4570 blend alone can exceed the strength of H$\upbeta$.

\begin{figure}
\centering 
\includegraphics[width=0.48\textwidth,angle=0]{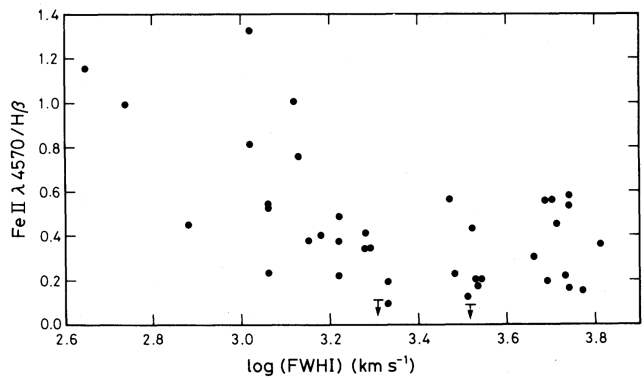}
\caption{Fe\,II/H$\upbeta$ versus the full width at half intensity (FWHI) of H$\upbeta$.  (Figure from \citealt{Gaskell85})}.
\end{figure}

\subsection{Strong object-to-object differences}

Differences in the equivalent widths or in the ratios of intensities of broad lines in AGNs are generally subtle.  Although Fe~II emission {\it can} be very strong, as in the example in Figure 1, this is not always the case.    It can be seen in Figure 2 that the ratio varies from greater than one to less that 0.1 -- more than an order of magnitude.  Such a large difference in broad-line ratios from AGN to AGN is rare. A satisfactory theory of Fe~II emission must not just explain the strength of Fe~II emission, but also why it is only strong in {\it some} objects. Not only does Fe~II emission differ in strength from object to object, but as we discuss in the following sections, the relative strength is correlated with other AGN properties.

\subsection{Correlation with line widths}

It can be seen from Figure 2 that the Fe~II $\uplambda$4570/H$\upbeta$ ratio is inversely correlated with the full width at half intensity (FWHI) of H$\upbeta$.  The ratio is strongest on average in narrow-line objects such as the example in Figure 1.

\subsection{Correlation with radio type}

\citet{Osterbrock77} and \citet{Grandi+Osterbrock78} found that Fe~II emission was often strong in radio-quiet Seyfert 1 galaxies, but not so strong in broad-line radio galaxies. \citet{Setti+Woltjer77} noted that among radio-loud AGNs there was a difference between extended, steep-radio-spectrum sources and compact, flat-spectrum sources. This was in the sense that the flat-spectrum sources had strong Fe~II emission like radio-quiet Seyfert 1s. The correlation of Fe~II emission with radio type was confirmed by \citet{Miley+Miller79}.

\subsection{Correlation with galaxy type}

Powerful radio-loud AGNs, which predominantly have weak Fe~II emission, are almost all in massive ellipticals and rarely in spirals \citep{Sandage72}. In contrast to this, radio-quiet AGNs, which  are have much stronger Fe~II emission on average, are overwhelmingly in spiral galaxies \cite{Adams77}.  There is thus a correlation of Fe~II emission with galaxy type and mass.

\subsection{Correlation with the narrow-line region}

\citet{Grandi+Osterbrock78} showed that AGNs with weaker Fe~II had stronger [O III].   \citet{Steiner81} classified AGNs into type A, which had strong Fe~II and type B, which had weak Fe~II. A similar classification was later made by \citet{Sulentic+00}.  Steiner showed that the two types had different NLR properties.  For type A, the ratio of [O III]/H$\upbeta$ was inversely correlated with the optical continuum luminosity.  For type B (the weak Fe~II emitters) there was no correlation and the [O III]/H$\upbeta$ ratio was systematically greater.

\citet{Boroson+Green92} performed a principle-component analysis of several properties of AGNs.  This combines separate correlations that had previously been identified.  Most of the variance is accounted for by Boroson and Green's eigenvector 1 (commonly referred to as just ``eigenvector 1'' or ``EV1'').  This is primarily the combination of the correlations of Fe~II/H$\upbeta$ with the FWHM of H$\upbeta$, and of Fe~II $\uplambda$4570/H$\upbeta$ with [O III]/H$\upbeta$.

\subsection{Correlation with soft X-ray excess}

The spectral energy distribution (SED) of a typical thermal AGN\footnote{A ``thermal AGN''  \citep{Antonucci12} is a high-accretion-rate AGN where the energy output is dominated by the thermal emission from the accretion disc; a ``non-thermal AGN'' is a low-accretion-rate AGN where the energy output is dominated by the mechanical energy in the jet.  These two types are referred to in the literature considering AGN feedback on galaxy evolution as ``quasar mode'' and ``radio mode'' respectively, but it is important to note that thermal AGNs can be either radio loud or radio quiet.} with the main sources of radiation for the main spectral regions labelled is shown in Figure 3.  This was derived by \citet{Gaskell+07} from reddening-corrected, multi-wavelength observations of NGC~5548.  As can be seen, the SED is dominated by the so-called ``Big Blue Bump'' (BBB), which is mostly due to the thermal emission from the accretion disc. The continuum shown in Figure 3 is similar to the widely-used continuum of \citet{Mathews+Ferland87}, derived independently mostly from photoionization considerations, and a similar continuum adopted by \citet{Korista+97}\footnote{The main difference is that the IR continuum is much higher in \citet{Mathews+Ferland87} because \citet{Gaskell+07} were deriving the continuum seen by the BLR. The {\it observed} IR, which comes from far outside the BLR, is higher relative to the higher-frequency continuum.}.  It also close to the continua with the largest BBB derived by \citet{Jin+12}, as shown on the left-hand side of their Figure 14\footnote{Note that \citet{Jin+12} mostly do not correct their objects for internal reddening, so their other continua have a too weak BBB.  An important physical constraint is that the BBB needs to be strong enough to produce the observed equivalent widths of the broad lines.}.

\begin{figure}
\centering 
\includegraphics[width=0.48\textwidth,angle=0]{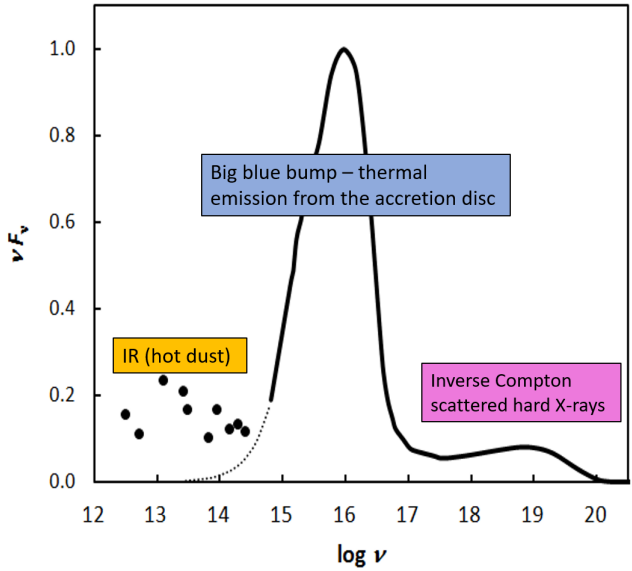}
\caption{The spectral energy distribution of a typical thermal AGN {\it as seen by the BLR}. Note that in a linear $\nu F_{\nu}$ vs. $\log \nu$ plot such as this, the area under curves is proportional to the energy.  Figure adapted from \citet{Gaskell08}.}
\end{figure}

In the hard X-ray region (energies of $2 - 10$ keV and greater) the flux per unit frequency, $F_{\nu}$, can be approximated as an $F_{\nu} \propto \nu^{-\alpha}$ power-law in frequency \citep{Mushotzky+80}.  For AGNs with a FWHM of H$\upbeta \gtrsim 4000$ km s$^{-1}$, the spectral index, $\alpha$, for energies greater than 2 keV is $\sim 0.75$ \citep{Zhou+Zhang10}.  If one extrapolates the hard X-ray power law down to lower energies (below 2 keV), the soft X-ray flux lies above the extrapolation of the hard X-ray power law.  This is referred to as the ``soft X-ray excess''.  This is the high-energy tail of the Big Blue Bump shown in Figure 3.  

\citet{Wilkes+87} found that the strength of the soft X-ray excess, as measured by the steepness of the X-ray spectral index from 0.1 to 3.5 keV, was correlated with the strength of optical Fe~II emission (see also \citealt{Shastri+93} for additional discussion). \citet{Puchnarewicz+92} found that AGNs with strong soft X-ray excesses had narrower Balmer lines.  Since strong Fe~II and narrow Balmer lines are one end of EV1, these two correlations mean that EV1 is correlated with the soft X-ray excess.  These correlations were confirmed by \citet{Boller+96} and \citet{Wang+96}.  \citet{Brandt+Boller99} showed that the correlation with EV1 itself was stronger than with the individual components making up EV1.  This is shown in Figure 4.

\begin{figure}
\centering 
\includegraphics[width=0.48\textwidth,angle=0]{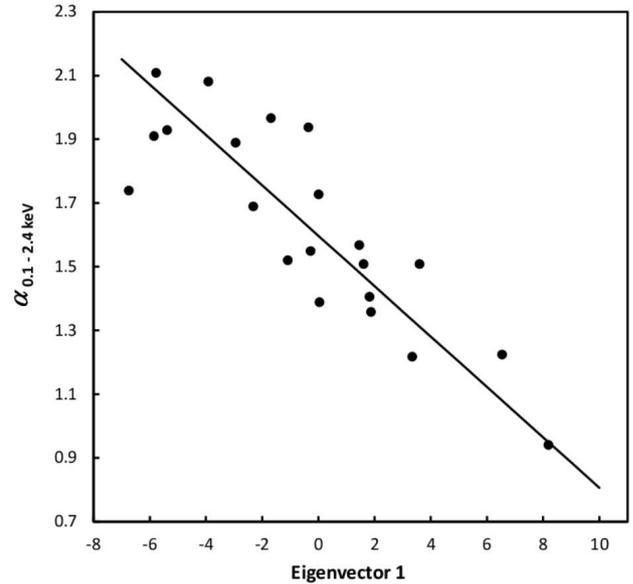}
\caption{The ROSAT soft X-ray, 0.1 -- 2.4 keV, spectral index, $\alpha$, versus EV1.  Data from \citet{Brandt+Boller99}.  The line is a least squares bisector fit.}.
\end{figure}

\section{How and where is Fe~II emission produced?}

To address the issue of the correlations of Fe~II emission with other AGN properties, it is essential to first answer a more fundamental question is: how is Fe~II emission produced?  \citet{Boller+96} remark ``The excitation mechanism of Fe~II lines is one of the outstanding problems of AGN research.”

It has long been established that the broad Balmer line strengths are correlated with the continuum flux as an AGN varies (\citealt{Lyutyi+Cherepashchuk72}; \citealt{Pronik+Chuvaev72};  \citealt{Cherepashchuk+Lyutyi73}).  This implies that the majority of the flux of the broad Balmer lines is produced by photoionization. The fluxes of Lyman $\alpha$ and C\,IV $\uplambda$1549 are also correlated with the continuum flux \citep{Penston+81}.   The same is true for the high-ionization He\,II $\uplambda$1640 line \citep{Antonucci+Cohen83}.  All those lines must therefore also be produced mostly by photoionization. 

Despite the evidence from variability that many broad lines must be predominantly produced by photoionization, doubts have long been raised about a photoionization origin of Fe~II.  In order to explain the strength of Fe~II emission, \citet{Collin-Souffrin+82} and \citet{Joly87} postulated the existence of low-temperature dense gas that was heated non-radiatively and emitting mainly the optical Fe~II.  \citet{Collin-Souffrin+88} proposed that Fe~II is produced by heating the outer parts of the accretion disc by hard X-rays with energies of up to hundreds of keV. \citet{Zheng+Keel91} suggested that the region producing optical Fe~II ``is not coextensive with the conventional broad-line region'' and but was consistent with coming from a region associated with a small-scale jet.  \citet{Joly91}, noting the correlation of Fe~II with radio properties suggested that the Fe~II region is closely associated with the jets responsible for the compact radio source. She proposed that the heating of the gas emitting Fe~II is due to internal shocks.  In a review of Fe~II emission, \citet{Collin+Joly00} state, `` the strengths of the Fe~II lines cannot be explained in the framework of photoionization models. A non-radiative heating, for instance due to shocks, with an overabundance of iron, can help to solve the problem. A comparison with other objects emitting intense Fe~II lines also favors the presence of strong outflows and shocks.''  \citet{Veron-Cetty+04} in making a detailed analysis of high-resolution, high signal-to-noise-ratio spectra of I~Zw~1 did not succeed in modelling the BLR with photoionization and  suggest that a non-radiative heating mechanism is needed to increase the temperature to provide the necessary additional excitation of the Fe~II lines.  \citet{Baldwin+04} also suggest that collisional excitation of warm dense gas was a possibility for explaining the ultraviolet Fe~II emission from $\uplambda\uplambda$2200--2800.

Another factor contributing to doubts about the origin of Fe~II emission was the difficulty in doing reverberation mapping of optical Fe~II. Perhaps Fe~II was not varying like the rest of the BLR? From monitoring of Akn 120 \citet{Kuehn+08} concluded that ``the optical Fe~II emission does {\it not} come from a photoionization-powered region similar in size to the H$\upbeta$-emitting region'' [emphasis by the authors] and they go on to speculate that there could be ``emission from a photoionized region several times larger than the H$\upbeta$ zone, or ... emission from gas heated by some other means, perhaps responding only indirectly to the continuum variations.''  Although they did find variations in Fe~II strength, early reverberation mapping campaigns such as those of \citet{Doroshenko+99}, \citet{Kollatschny+00}, \citet{Wang+05} and \citet{Shapovalova+12} failed to get useful time delays.  

\section{Is Fe~II emission caused by photoionization?}

Despite the doubts just enumerated, we can now readily and strongly answer in the affirmative that Fe~II {\it is} caused by photoionization because, as we will discuss below, Fe~II reverberation mapping has now been successfully carried out over the past decade for a number of AGNs.  In Figure 5 we show superimposed shifted and normalized light curves for the $\uplambda$5100 (rest frame) continuum, H$\upbeta$, and Fe~II for 3C 273 over the ten-year period (2008 - 2018).  We have taken fluxes from \citet{Zhang+19}. To reduce the scatter we have averaged every nine nights.  This enables us to know the error in the mean for each average point.  We have scaled the three curves to the same mean (zero) and the same variance.  The RMS fractional variability amplitudes of H$\upbeta$ and Fe~II after allowance for measuring errors are 5.2\% and 3.8\% respectively (see Table 2 of \citealt{Zhang+19}).  This makes the RMS fraction Fe~II variability amplitude $\sim 73$\% of the H$\upbeta$ amplitude.    Fe~II thus appears to be only slightly less variable than H$\upbeta$. We will argue below that the responses of H$\upbeta$ and Fe~II to continuum variability are actually the same. The H$\upbeta$ and Fe~II points have been shifted back relative to the continuum points by 270 and 450 days respectively.

\begin{figure}
\centering 
\includegraphics[height=0.39\textwidth,angle=0]{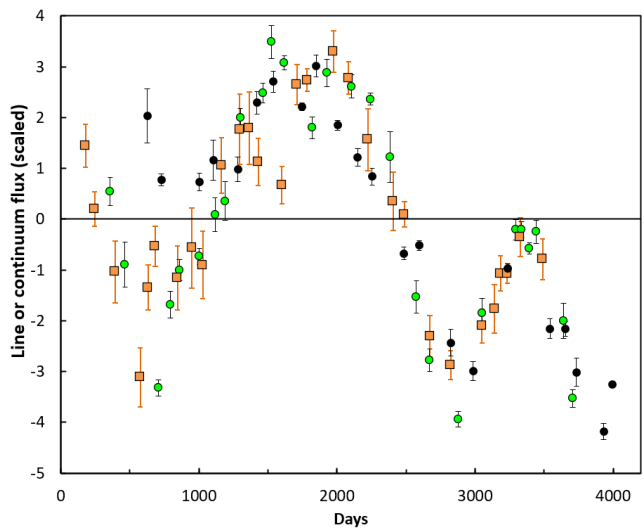}
\caption{Comparison of light curves for the optical continuum (black circles), H$\upbeta$ (green circles) and Fe~II (brown squares) from 11 years of monitoring of 3C 273 from 2008 to 2018.  Each point is the average of nine nights and the $\pm 1$$\sigma$ error bars show the errors in each mean.  The three curves have been normalized to the same standard deviation and zero mean.  The time axis is in observed relative times after shifting (see text.)}
\end{figure}

What is obvious from a casual glance at Figure 5 is that, after the shifts in time have been applied, H$\upbeta$ and Fe~II track each other very well.  Thus, whatever is causing H$\upbeta$ to vary is also similarly causing Fe~II to vary as well.  One can also see that both H$\upbeta$ and Fe~II track the broad features of the continuum variability.  Thus there can be little doubt that {the variability of both lines is due to photoionization.}  However, closer inspection of Figure 5 reveals that there are some anomalies. We will discuss these below.

\section{The structure of the BLR: the Gaskell, Klimek \& Nazarova model}

Given that there can be little doubt now that Fe~II emission is produced by photoionization, and that it has long been recognized that other BLR lines are produced by photoionization, we can ask: what is the expected relationship of the region emitting Fe~II to the region(s) emitting the other lines? 

The probable structure of the BLR is reviewed in \citet{Gaskell09}.   From consideration of the energetics of NGC~5548, \citet{Gaskell+07} (GKN) deduced that the low-ionization component of the broad-line region of an AGN has a flattened distribution (see also \citealt{Gaskell09}).  The line strengths require a relatively high covering factor, but the lack of Lyman-edge absorption in AGNs requires that there be a ''hole at the pole'' \citep{Maiolino+01}. The motions of the clouds are highly turbulent and they are co-rotating with the accretion disc.  This distribution correctly produces observed BLR line profiles (see Figure 2 of \citealt{Gaskell11}) and the distribution has been well confirmed by subsequent high-quality reverberation mapping (see, for example, \citealt{Pancoast+14}) 

\begin{figure}
\centering 
\includegraphics[height=0.39\textwidth,angle=0]{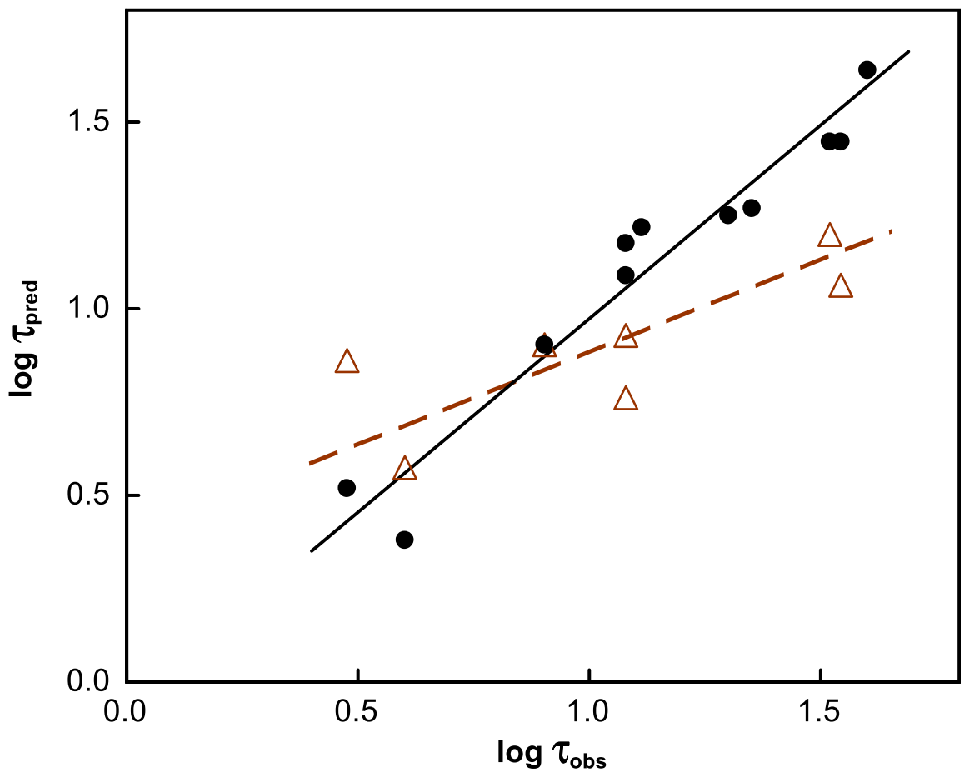}
\caption{Predicted reverberation mapping time lags,  $\tau_{pred}$, versus observed lags, $\tau_{obs}$ for lines of different ionization in NGC~5548.  The solid circles are for the predictions of GKN (see their Table 6).  The open triangles are the LOC model predictions of \citet{Korista+Goad00} except for the left-most triangle, which is the He~II prediction of \citet{Bottorff+02}. Observed lags are as reported by \citet{Clavel+91}, \citet{Peterson+91}, \citet{Krolik+91}, and \citet{Bottorff+02}.  The lines are least-squares bisector fits.  (Figure from \citealt{Gaskell09}).}
\end{figure}

An important implication of this flattened distribution is that {\it the BLR will completely cover the central source when viewed from far away near the equatorial plane}. The radiation experienced by the outer BLR and by dust near the equatorial plane has thus been heavily absorbed by the inner BLR.  

The self-shielding has important consequences for where lines are emitted from. The nearly universal assumption in using photoionization codes such as {\it Cloudy} \citep{Ferland+98} for modelling emission from the BLR is that there are large, optically-thick clouds, each with an unobstructed view of the source of ionizing radiation. A first approximation \citep{Davidson72} was to assume a ``typical'' standard BLR cloud.  Davidson showed that an improved fit to observed spectra could be obtained by having two sets of clouds with different ionization parameters.  An even greater improvement in explaining total line strengths was introducing a population of clouds with a whole range of densities and ionization parameters -- the ``locally-optimal clouds'' (LOC) model of \citet{Baldwin+95}. Although they average a large range of densities and ionization parameters, the LOC {\it Cloudy} models retain the assumption that the individual clouds are optically thick, so that each cloud emits lines with a range of ionization, and that they have an unobstructed view of the source of ionizing radiation.

Noting that the geometry of the BLR required the system of BLR clouds to be self-shielding, \citet{Gaskell+07} proposed that rather than having large, individually optically thick BLR clouds, all directly exposed to the ionization radiation, instead, the more distant clouds are shielded from much of the direct radiation by the closer-in clouds.  This means that the ionization structure of a single optically-thick cloud in {\it Cloudy} spread over a fraction of an astronomical unit is now the ionization structure of an ensemble of clouds that, in the most luminous AGNs, can span a light year.  This sort of model is a return to a model that was considered early on in BLR modelling by \citet{MacAlpine72}.\footnote{It is interesting to compare the ionization structure of a single cloud as shown in the right hand side of Figure 2 of \citet{Davidson72} with that the ensemble of clouds shown of Figure 1 of \citet{MacAlpine72}. The curves are qualitatively similar, but the horizontal axis of the former is labelled in units of $10^{13}$cm (about an AU), while the latter is labelled in parsecs.}  An important success of the GKN self-shielding model is that it predicts a high degree of ionization stratification.  If we assume that the ionization parameter is roughly constant with radius (as is observed to be the case even out as far as the narrow-line region -- see \citealt{Stern+14}) and that  the BLR filling factor is roughly constant with radius, then the model readily predicts the effective radii of regions emitting lines of different ionization.  Optically-thick cloud models such as the LOC model predict less variation of effective radius with ionization than is observed (see Figure 6).

\section{Photoionization modelling of Fe~II emission}

\citet{Gaskell+07} did not predict the effective radius of Fe~II emission because no reverberation-mapping lags were available at that time.  We have therefore run new photoionization models to estimate the effective radius of Fe~II emission.  

\subsection{Model parameters}

Models were run using release 17.10 of {\it Cloudy} \citep{Ferland+17}.  For consistency with previous work we used the default standard \citet{Mathews+Ferland87} continuum, but, as we have noted, this is very similar to the continuum derived for NGC~5548 by \citet{Gaskell+07} and the continuum used by \citet{Korista+97}. We ran models with a constant hydrogen density, $n_H = 10^{11}$ cm$^{-3}$, and solar abundances. Models were allowed to run to a column density of $10^{25}$ cm$^{-1}$ to avoid effects of artificially imposing matter bounds on the distribution of gas. For a fixed continuum shape, the ionization structure of a nebula depends primarily on the ionization parameter, $U_1$, the ratio of density of ionizing photons with energies greater than one Rydberg to the gas density. We ran models over the range $-2.5 < \log U_1 < +0.5$ to investigate the effect of variability. To measure optical Fe~II emission we used the $\uplambda$4584 and $\uplambda$5317 lines as calculated by {\it Cloudy} as representatives of the $\uplambda$4570 and $\uplambda$5300 blends. The calculated fluxes of the two individual lines were well correlated with each other and with the majority of other optical lines.  As well as giving the results of calculating the expected radius of Fe~II emission, we also briefly mention a couple of other aspects of Fe~II emission.

\begin{figure}
\centering 
\includegraphics[height=0.46\textwidth,angle=0]{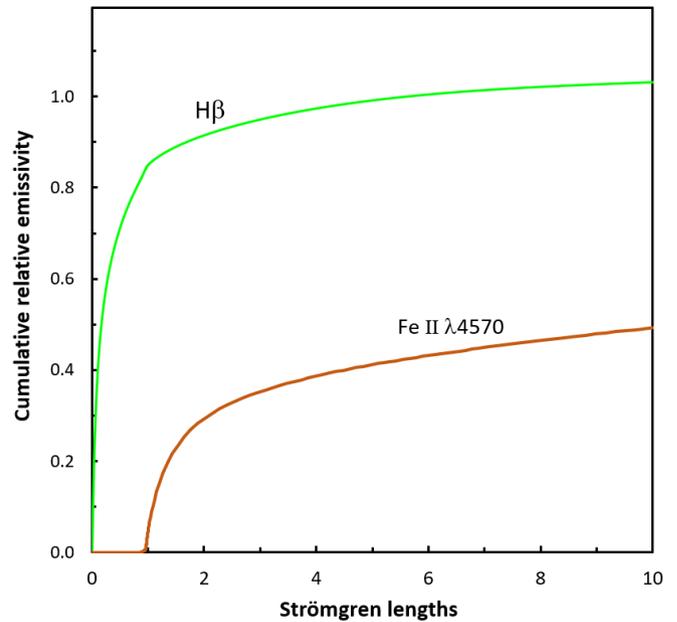}
\caption{Cumulative relative line fluxes of H$\upbeta$ and Fe~II $\uplambda4570$ as a function of the hydrogen Str\"{o}mgren length.}
\end{figure}

\begin{figure}
\centering 
\includegraphics[height=0.47\textwidth,angle=0]{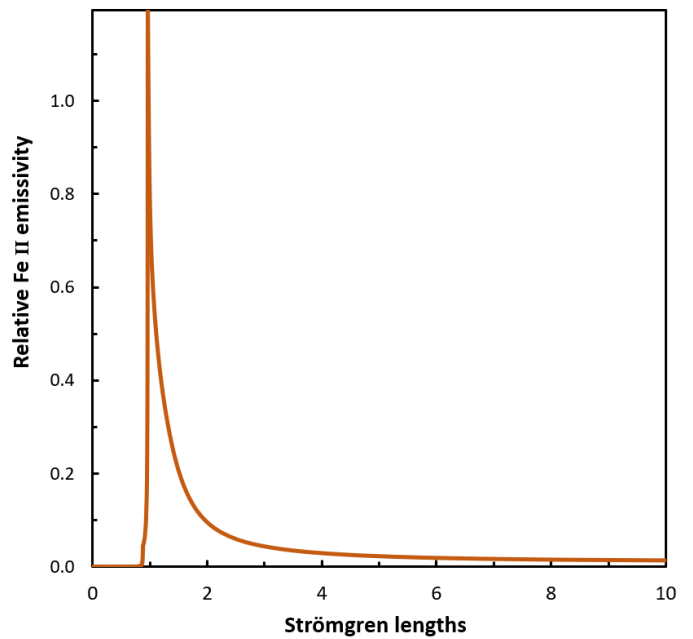}
\caption{Relative emissivity of Fe~II $\uplambda$45470 (in arbitrary units) as a function of the hydrogen Str\"{o}mgren length.}
\end{figure}

\subsection{Strength of optical Fe~II emission}

Figure 7 shows an example of the cumulative relative line intensities for H$\beta$ and the Fe~II $\uplambda$4570 blend. We give these as a function of the hydrogen Str\"{o}mgren length (taken as the inflection point where the H$\upbeta$ emission drops off).  Figure 8 shows the relative emissivity of Fe~II $\uplambda$4570.  With an ionization potential of only 16.18 eV, Fe$^+$ is readily ionized.  It only appears after much of the ionizing radiation starts to be absorbed as the fraction of neutral hydrogen increases.  Thus, as can be seen, Fe~II emission begins as hydrogen become predominantly neutral in the partially-ionized zone.  The strongest Fe~II emission is from just beyond Str\"{o}mgren length where the temperature is still high.  As the column density increases further beyond a couple of Str\"{o}mgren lengths, both H$\upbeta$ and Fe~II emission continue to increase slowly, Fe~II somewhat faster than H$\upbeta$. In a real AGN there will be no emission from clouds beyond a few Str\"{o}mgren lengths if the GKN model is correct because one gets to the surrounding dust.

The total flux of lines in the wavelength range $\uplambda$4400 to $\uplambda$4800 that make up the $F_{4570}$ blend used in the ratio $R_{\mathrm {FeII}} = F_{4570}/F_{{\mathrm H}\upbeta}$ as a measure of the relative strength of Fe~II emission, give $R_{\mathrm {FeII}} \thickapprox 0.35$ for our models with solar abundance.  This is comfortably in the observed range (see Figure 2).  $R_{\mathrm {FeII}}$ increases with the gas-phase [Fe/H] abundance (see Sections 15 \& 16) and since [Fe/H] is expected to be two or three times solar, or perhaps even a little more (see Section 16), the models have no difficulty covering the observed range in Figure 2.

\subsection{Responsivities of Fe~II and H$\upbeta$}

It was initially widely believed that Fe~II emission in AGNs is driven by hard X-ray emission penetrating far into the warm partially-ionized gas.  Since 2--10 keV X-ray emission is not well correlated with optical continuum emission (see \citealt{Peterson+00} and \citealt{Maoz+02}), this predicts that the variability of Fe~II emission could be quite different from that of H$\upbeta$.  Also, if Fe~II emission is primarily produced by hard X-ray emission, this predicts that a flatter optical to X-ray spectral index, $\alpha_{\mathrm{ox}}$ would give stronger Fe~II emission.  In fact, quite the opposite is found: a steeper $\alpha_{\mathrm{ox}}$ correlates with a {\it larger} $R_{{\mathrm {FeII}}}$ (e.g., \citealt{Wang+96}). 

From Figure 8 it can be seen that while there is Fe~II emission from much higher column densities (and the models show it continuing up to hundreds of Str\"{o}mgren lengths), most Fe~II emission originates {\it within a couple Str\"{o}mgren lengths}. As noted, in the GKN model the BLR cannot extend more than a few Str\"{o}mgren lengths in a real AGN because of the surrounding dust.  At the optical depths at which most of the Fe~II is produced, the heating and ionization are due to lower energy photons rather than hard X-rays. We find that the relative responses of Fe~II and H$\upbeta$ emission are similar as the continuum level changed by orders of magnitude.

\subsection{Predicted effective radii for Fe~II}

Our models give an emissivity-weighted radius of Fe~II that 2.1 times that of H$\upbeta$.  Reverberation mapping measures the {\it responsivity}-weighted radii of lines rather that the emissivity-weighted radii, so we calculated the relative responsivity-weighted radii for H$\upbeta$ and Fe~II from the differences in emissivity as a function of radius at different flux levels.  The response of H$\upbeta$ with increasing ionizing flux is positive at all radii.  However, for Fe~II emission, which arises almost entirely beyond one Str\"{o}mgren length, the responsivity goes negative just beyond the initial Str\"{o}mgren length as the length increases with increasing ionizing flux. This is because Fe$^+$ is lost to Fe$^{++}$.  We calculated the ratio of responsivity-weighted radii for different hydrogen column densities (i.e., matter-bounded clouds) and found that, so long as the column is large enough that in the high state there is gas to around at least 1.5 times the Str\"{o}mgren length, then the ratio of radii is constant at $\sim 2.2$.  If one continues to very high column densities (many times the Str\"{o}mgren length), Fe~II emission continues to steadily increase more than H$\upbeta$ emission, but the emission from both lines becomes very small because of the falling temperature, so there is little increase in the ratio of radii.  Again, as noted, extremely high column densities (many tens or hundreds of Str\"{o}mgren lengths) will not occur in real AGNs with the GKN model. In summary then, the GKN model predicts that lag of Fe~II measured by reverberation mapping should be close to twice that of H$\upbeta$.  This radius is also the same radius the O~I is emitted at (the top-right point in Figure 6), and only slightly ($\sim 10$\%) greater than the radius Mg~II is emitted at.

\section{Line widths}

The first indication of the location of the Fe~II emitting region came from the widths of the Fe~II lines.  From his detailed analysis of individual Fe~II lines in I~Zw~1 \citet{Phillips78a} concluded that Fe~II had a full width at half maximum (FWHM) of between about 75\% and 100\% of that of H$\upbeta$. The narrower widths of Fe~II lines compared with H$\upbeta$ can be appreciated in Figure 9.  \citet{Shuder82} argued by comparing broad line profiles in AGNs that the lines with the widest profiles are closest to the central source of ionizing radiation.  \citet{Gaskell+Sparke86}, from reverberation mapping of lines of different ionizations, found larger sizes for regions emitting lower ionization lines.  This was confirmed by further observations  \citep{Koratkar+Gaskell89,Koratkar+Gaskell91}. \citet{Gaskell88} and \citet{Koratkar+Gaskell89} showed from velocity-resolved reverberation mapping that the gas emitting the bulk of both the high- and low-ionization lines is gravitationally bound.  Furthermore, \citet{Krolik+91} showed that the widths of broad lines are correlated with the sizes of the emitting regions given by reverberation mapping, just as is expected if the motions are gravitationally dominated (see their Figure 4). We can thus expect the width of Fe~II lines to be an indicator of how far away from the black hole they are produced.

\begin{figure}
\centering 
\includegraphics[height=0.47
\textwidth,angle=0]{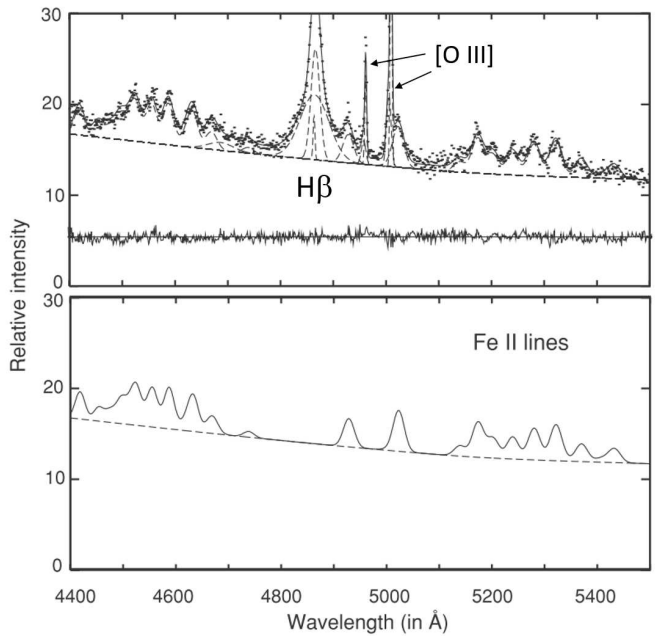}
\caption{Example of a fit to a spectrum with residuals.  The top panel shows the spectrum of SDSS J020039.15 - 084554.9, the fit continuum, NLR [O III] and H$\upbeta$, Fe~II, broad H$\upbeta$, and He~II $\uplambda$4686. The bottom panel shows just the Fe~II model. (Figure adapted by permission from \citealt{Kovacevic+10})}
\end{figure}

\citet{Hu+08} give full width at half maximum (FWHM) measurements of a very large sample of SDSS AGNs. Their measurements show that Fe~II has a FWHM of 71\% of H$\upbeta.$\footnote{Note that the Fe~II velocity shifts claimed by \citet{Hu+08} are not confirmed by other analyses.  See \citet{Sulentic+12} and \citet{Bon+18} for discussion.}  This is identical to the relative widths \citet{Marinello+16} find for O I, Ca II, and infra-red Fe emission compared with P$\upbeta$.  From the relationship between the effective radius and FWHM \citep{Krolik+91}, the relative widths of the Fe~II lines imply that most of the Fe~II emission comes from twice as far out as the bulk of the H$\upbeta$ emission.

\section{Reverberation mapping}

In principle, reverberation mapping is a more direct way of getting the radius of the Fe~II region.   In practice, reverberation mapping of optical Fe~II emission has unfortunately proved to be considerably harder than reverberation mapping of H$\upbeta$ and high-ionization lines. While this could be due to Fe~II responding to a different continuum to H$\upbeta$, we believe that there are other explanations of the difficulty of Fe~II reverberation mapping. There are two main problems.

\subsection{Problem number one: Fe~II variability is reduced and smeared out because of the larger region size} 

We believe that the main problem is simply the larger light travel time predicted for the Fe~II emitting region, a factor of two larger than for H$\upbeta$ (see above).  We have investigate the effects of this with simple simulations. As suggested by \citet{Gaskell+Peterson87}, AGN UV and optical continuum variability can be modelled as a damped random walk (DRW) with a characteristic damping time, $\tau_{\mathrm{damp}}$. The effective radius of the BLR emitting H$\upbeta$ is of the order of a few times the characteristic optical continuum variability timescale (see Figure 2 of \citealt{Dibai+Lyutyi84}).  For our modelling we therefore took $\tau_{\mathrm{damp}} = 10$ days for our simulated DRW continua and convolved them with boxcar response functions with lags of 50 days and 100 days to generate artificial line light curves for H$\upbeta$ and Fe~II respectively. We show two examples in Figure 10.  Our response functions for each line had a half width of half the lag.  The shape of the response function has a negligible effect on our simulations. 

\begin{figure}
\centering 
\includegraphics[height=0.52\textwidth,angle=0]{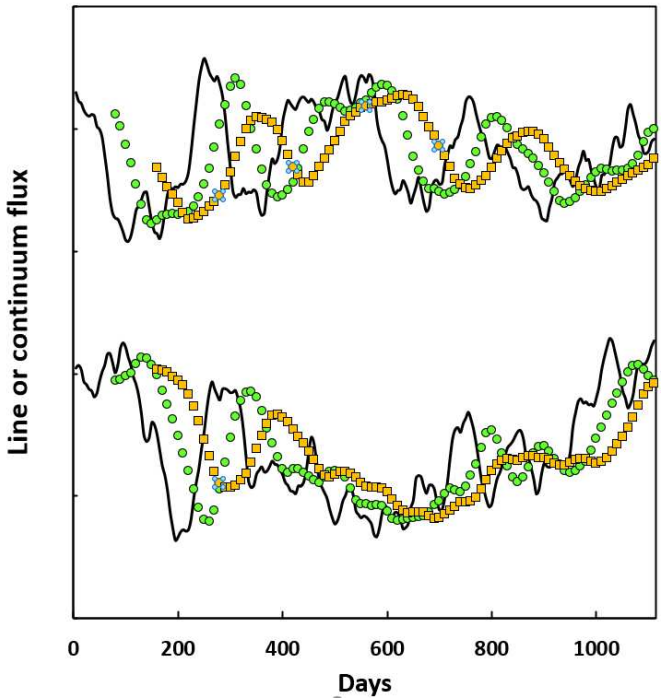}
\caption{Two sets of simulated continua, H$\upbeta$ and Fe~II light curves. The continuum variations, shown as black curves, are modelled as a damped random walk as per \citet{Gaskell+Peterson87} and the line curves have calculated assuming $F_{line} \propto F_{cont}$ with lags of 50 days and 100 days respectively for H$\upbeta$ (green circles) and Fe~II (brown squares). The response functions have been taken to be boxcar functions with half widths equal to half the lag. The upper set of curves has been arbitrarily offset to avoid confusion in plotting.}
\end{figure}

The simulations shown in Figure 10 illustrate that Fe~II emission can often be less variable than H$\upbeta$ because of the washing out of variability.  For example, in the lower simulation shown, Fe~II is nearly constant between day 600 and day 800, despite the continuum variability during this interval.  It does not follow the continuum variability as clearly as H$\upbeta$ does around days 700 to 1000.  Even when there is clear Fe~II variability, as around days 200 to 400 in both simulations, the amplitude of Fe~II variation is less than for H$\upbeta$.

From our simulations we can calculate how much the variability amplitude is reduced.  We find that the average variability amplitude of Fe~II deduced from our synthetic light curves is reduced on average to $ \thickapprox 83$\% of that of H$\upbeta$, even though we put in identical responses to the continuum changes. The median observed ratio of the amplitudes of Fe~II variability to H$\upbeta$ variability as reported by \citet{Barth+13}, \citet{Hu+15}, \citet{Zhang+19}, \citet{Hu+20} and \citet{Lu+21} is also 83\%.    Thus, after allowance for the smearing by a broader Fe~II response function, the responses of Fe~II and H$\upbeta$ are consistent with being the same.

The simulations shown in Figure 10 are noise-free and perfectly sampled.  Real AGN monitoring data are neither.  If one only has every tenth point in Figure 10, it becomes hard to detect features in the Fe~II light curve even without adding noise.  Because the Fe~II emission features are broad (see Figure 1), they are not as easy to measure as a well-defined lines like H$\upbeta$.  They are thus subject to larger random and systematic errors (see \citealt{Bon+18}).  It can be seen that the error bars of the mean fluxes in Figure 5 are indeed larger for Fe~II than for H$\upbeta$.

\subsection{Problem number two: Narrow-line Seyfert 1s are less variable than broad-line AGNs}

Although the total emission in Fe~II line is strong, it is spread out over a wide spectral range (see, for example, Figure 1).  This makes it hard to measure when the lines are broad.  For successful reverberation mapping, one ideally needs both strong and easy-to-measure features and a highly-variable AGN. Because Fe~II is easy to measure, NLS1s have been attractive targets for reverberation mapping. Unfortunately, NLS1s are observed to be significantly less variable than non-NLS1s (\citealt{Klimek+04}; \citealt{Ai+10}). The difference in real variability is even greater because the host galaxy starlight dilution, which reduces apparent variability, is much greater on average in low-Eddington-ratio AGNs (i.e., NLS1s) than in high-Eddington-ratio ones \citep{Gaskell+Kormendy09}.  

As a consequence of a small amplitude of optical and UV variability a two-year campaign with 50 days of intensive X-ray, UV and optical monitoring of the NLS1 Akn 564 failed to get reliable lags for {\it any} optical lines \citep{Shemmer+01}. \citet{Doroshenko+06} subsequently concluded on the basis of 17 years of photometry that ``Ark 564 is one of the least variable Seyfert galaxies in the optical.''  The failure to measure lags of emission lines in Akn 564 was, naturally, a deterrent to mounting similar NLS1 multi-wavelength monitoring campaigns.

The monitoring of Akn 564 revealed an additional problem: not only was the {\it amplitude} of UV and optical variability low, but the variability was also {\it rapid} (see Figure 3 of \citealt{Gaskell+Klimek03}).  Rapid continuum variability makes reverberation mapping of large emission line regions difficult.\footnote{On the other hand, the rapid continuum variability of Akn 564 did facilitate the measurement of wavelength-dependent lags in the continuum.}
 
\begin{table*}
\centering															
\caption{Reverberation mapping lags for H$\upbeta$ and Fe~II}%
\begin{tabular}{l  c c c c  c}										
\toprule
Object	&	Log $\tau_{{\mathrm H}\upbeta}$	 		&	Log $\tau_{\mathrm{FeII}}$			&	Log $\tau_{\mathrm{FeII}}/\tau_{\mathrm{H}\upbeta}$			&	$\tau_{\mathrm{ FeII}}/\tau_{\mathrm{H}\upbeta}$	&		Reference		\\
\midrule
Mrk 335	&	0.94	$\pm$	0.06	&	1.43	$\pm$	0.04	&	0.49	$\pm$	0.07	&	3.1	&	a	\\
3C 273	&	2.23	$\pm$	0.03	&	2.57	$\pm$	0.08	&	0.34	$\pm$	0.08	&	2.2	&	b	\\
Mrk 817	&	1.45	$\pm$	0.03	&	1.71	$\pm$	0.08	&	0.26	$\pm$	0.09	&	1.8	&	c	\\
Mrk 1511	&	0.77	$\pm$	0.07	&	0.94	$\pm$	0.07	&	0.17	$\pm$	0.09	&	1.5	&	d	\\
NGC 4593	&	0.64	$\pm$	0.10	&	0.92	$\pm$	0.07	&	0.29	$\pm$	0.13	&	1.9	&	d	\\
PG 2130+099	&	1.35	$\pm$	0.06	&	1.55	$\pm$	0.11	&	0.19	$\pm$	0.13	&	1.6	&	e	\\
Mrk 142	&	0.90	$\pm$	0.06	&	0.88	$\pm$	0.11	&	-0.02	$\pm$	0.13	&	1.0	&	a	\\
	&				&				&				&		&		\\
Mrk 486	&	1.37	$\pm$	0.09	&	1.24	$\pm$	0.11	&	-0.14	$\pm$	0.14	&	0.7	&	a	\\
Mrk 382	&	0.88	$\pm$	0.14	&	1.38	$\pm$	0.11	&	0.50	$\pm$	0.18	&	3.2	&	a	\\
Mrk 1044	&	1.02	$\pm$	0.12	&	1.14	$\pm$	0.14	&	0.12	$\pm$	0.18	&	1.3	&	a	\\
MCG+06--26--012	&	1.38	$\pm$	0.11	&	1.35	$\pm$	0.15	&	-0.03	$\pm$	0.19	&	0.9	&	a	\\
IRAS F12397+3333	&	0.99	$\pm$	0.14	&	1.03	$\pm$	0.15	&	0.04	$\pm$	0.21	&	1.1	&	a	\\
PG 1700+518	&	2.22	$\pm$	0.16	&	2.58	$\pm$	0.16	&	0.18	$\pm$	0.22	&	1.5	&	f	\\
Mrk 493	&	1.06	$\pm$	0.08	&	1.08	$\pm$	0.23	&	0.01	$\pm$	0.24	&	1.0	&	a	\\
IRAS 04416+1215	&	1.12	$\pm$	0.08	&	1.10	$\pm$	0.35	&	-0.02	$\pm$	0.39	&	0.9	&	a	\\
PG 0026+120	&	2.22	$\pm$	0.08	&	2.18	$\pm$	0.41	&	-0.04	$\pm$	0.42	&	0.9	&	f	\\
\bottomrule																\\																	
\end{tabular}															\begin{tablenotes}
References: (a)  \citet{Hu+15}, (b) \citet{Zhang+19}, (c) \citet{Lu+21}, (d) \citet{Barth+13}, (e) \citet{Hu+20}, (f) \citet{Chelouche+14}. Lags are in days.  
\end{tablenotes}													\end{table*}																	

\subsection{Reverberation-mapping results}

Thanks to large observing campaigns taking care to minimize of errors in data taking and analysis (see, for example, the discussion in \citealt{Hu+15}), there has now been success over the last decade in obtaining reliable Fe~II lags from reverberation mapping, despite the aforementioned difficulties. These results unambiguously support Fe~II being produced by photoionization.  However, contrary to the predictions from the GKN model and Fe~II line widths, \citet{Hu+15} concluded from their reverberation mapping of nine NLS1s that ``the results support the idea that Fe~II emission lines originate in photoionized gas, which, for the majority of the newly-reported objects, is indistinguishable from the H$\upbeta$-emitting gas.''  Similarly \citet{Chelouche+14} concluded that reverberation mapping of Fe~II implied ``an emission-region size that is comparable to and at most twice that of the H$\upbeta$ line.'' \citet{Chelouche+14} and \citet{Hu+15} are thus concluding that the ratio of effective radii of emission of Fe~II and H$\upbeta$,  $\tau_{\mathrm{FeII}}/\tau_{\mathrm{H}\upbeta} \thickapprox 1$.  

Since similar lags are in conflict with the predictions of both the GKN model and the line widths that $\tau_{\mathrm{FeII}}/\tau_{\mathrm{H}\upbeta} \thickapprox 2$, we did our own determination of the lags from the reported observations to see if there were any problems.  We failed to find any problems and our determinations of lags agree within the errors with those previously reported.  However, we noticed that $\tau_{\mathrm{FeII}}/\tau_{\mathrm{H}\upbeta}$ was a function of the reported measuring errors.  

In Table 1, we give a compilation of the best reported H$\upbeta$ and Fe~II lags.  We have only include AGNs for which the error in log $\tau_{\mathrm{FeII}}/\tau_{\mathrm{H}\upbeta}$ is less than 0.5 (i.e, when the ratio is uncertain by less than a factor of three).  AGNs are listed in order of decreasing accuracy of the determination of the $\tau_{\mathrm{ FeII}}/\tau_{\mathrm{H}\upbeta}$ ratio, as given by the reported errors (column 4).   The lags are as given in the papers referenced. Errors are the means of the reported errors.  Since we are concerned with ratios we give the logarithms of the quantities.  As expected, the median errors in $\tau_{\mathrm{FeII}}$ are greater than the errors in $\tau_{\mathrm{H}\upbeta}$ by 50\%.  

For the objects in the upper half of the table (the first seven AGNs) it is possible to detect the expected ratio of $\tau_{\mathrm{ FeII}}/\tau_{\mathrm{H}\upbeta} = 2$ at better than $2.5\upsigma$. If we restrict ourselves to just these seven, the mean $\log \tau_{\mathrm{FeII}}/\tau_{\mathrm{H}\upbeta}$ is $0.23 \pm 0.06$, giving a $\tau_{\mathrm{FeII}}/\tau_{\mathrm{H}\upbeta} = 1.68^{+0.27}_{-0.23}$.  For the top four AGNs $\tau_{\mathrm{ FeII}}/\tau_{\mathrm{H}\upbeta} = 2$ can be detected at better than $3\upsigma$.  For these four AGNs with the best-determined lags, the mean $\log \tau_{\mathrm{FeII}}/\tau_{\mathrm{H}\upbeta}$ is $0.31 \pm 0.07$, giving $\tau_{\mathrm{FeII}}/\tau_{\mathrm{H}\upbeta} = 2.08^{+0.35}_{-0.30}$.  We therefore conclude that the best reverberation mapping gives an effective radius of Fe~II emission consistent with twice that of H$\upbeta$, in agreement with the predictions of the GKN model and the relative line widths.

\subsection{Biases in sizes from reverberation mapping}

For the nine AGNs studied by \citet{Hu+15}, the median $\tau_{\mathrm{FeII}}/\tau_{\mathrm{H}\upbeta}$ is 1.0 in agreement with their conclusion that the gas emitting Fe~II is indistinguishable from that emitting H$\upbeta$.  We believe that this is related to the accuracy of the determination of $\tau_{\mathrm{FeII}}/\tau_{\mathrm{H}\upbeta}$.  Seven of the nine AGNs studied by \citet{Hu+15} fall in the lower part of Table 1 and an additional AGN for which they did not successfully measure a lag would also belong in the lower part.  Only two of their sample are in the upper half.  The median $\tau_{\mathrm{FeII}}/\tau_{\mathrm{H}\upbeta}$ ratio for the most accurate ratios (the top seven AGNs in Table 1) is 1.8, while it is 1.0 for the poorer quality measurements.  A Mann-Whitney test gives a probability $p = 0.04$ for the null hypthesis that $\tau_{\mathrm{FeII}}/\tau_{\mathrm{H}\upbeta}$ is the same for both the higher- and lower-quality data.

We believe that this significant difference in $\tau_{\mathrm{FeII}}/\tau_{\mathrm{H}\upbeta}$ could be a consequence of biases identified by \citet{Welsh99}.  He showed through simulations that reverberation give systematically too small lags (i) when the observing campaign is short compared with the duration of the lag and (ii) when the measurement accuracy is low. These combination of these effects is illustrated in Figures 8 and 9 of \citet{Welsh99}. Low signal-to-noise has a stronger effect causing underestimation of lags when the duration of the observing campaign is short compared with the lag. These biases will lower $\tau_{\mathrm{FeII}}/\tau_{\mathrm{H}\upbeta}$ because the lags of Fe~II will be affected more than those of H$\upbeta$ for two reasons.  The first is that, because the lag of Fe~II is twice that of H$\upbeta$, the ratio of the length of an observing season to the time lag is less for Fe~II.  The second reason is that the signal-to-noise ratio for Fe~II flux measurements is about half that of the H$\upbeta$ flux measurements.  For the observing campaigns in Table 1, the median ratio of the campaign duration to the expected Fe~II lag is only a factor of six.  For the three worst cases (Mrk 486, MCG+06-26-012 and Mrk 496) it is only a factor of two.  All three have low values of $\tau_{\mathrm{FeII}}/\tau_{\mathrm{H}\upbeta}$.

\subsection{``Anomalies'' in reverberation mapping responses}

In Figure 5, the fluxes of H$\upbeta$, Fe~II and the optical continuum of 3C~273 agree well for the last five years of the monitoring (after day 2500 in the figure), but the continuum is much higher for the first 1000 days (three years) than one would expect from the line fluxes.  Given the realistic error bars, this is highly significant.  Another anomaly is that the continuum falls off more rapidly around day 2000 than implied by the line fluxes.  This is again very significant given the large number of independent observations. \citet{Gaskell+21} show that such anomalies are common in all type-1 thermal AGNs; agreement between observed and predicted H$\upbeta$ fluxes in AGNs is rarely perfect.  \citet{Gaskell+21} identify anomalies of two types: H$\upbeta$ being higher or lower than what one would expect from the optical continuum (as is the case here for 3C 273), and lags changing on the timescale of variability.  Possible causes of these anomalies are discussed in \citet{Gaskell+21}.  For the best-studied AGN, NGC~5548, the anomalies are on a timescale of about twice the H$\upbeta$ lag.  Since this is the timescale of the Fe~II lag, these anomalies introduce an additional source of error in trying to use reverberation mapping to determine the size of the region emitting Fe~II.

\section{The Fe~II emitting region in context}

\subsection{Relationship to other low-ionization lines}

As discussed, the evidence points to the effective radius of emission of Fe~II being twice that of H$\upbeta$.  This is the same size as the region emitting O~I as calculated by GKN and as determined by reverberation mapping of NGC~5548 \citep{Krolik+91}. Because NGC~5548 has weak Fe~II emission, it is not a good candidate for Fe~II reverberation mapping, but if NGC~5548 gives similar results as for AGNs for which $\tau_{{\mathrm{FeII}}}$ has been determined, then the point for Fe~II in Figure 6 would lie on top of the O I point, which is the top right point in the plot. Mg~II also lies in the upper right of the diagram.  Reverberation mapping of Ca~II has not been carried out, but we would expect it too to have a similar lag as Fe~II because \citet{Marinello+16} find for O I, Ca II, and Fe~II to all have similar line widths.  This is not surprising.  \cite{Wampler+Oke67} noted the Mg II and Fe~II should arise in similar gas.  The same is true for O I and Ca II.  This is because O$^\mathrm{o}$, Mg$^+$, Ca$^+$ and Fe$^+$ have ionization potential of 13.6, 15.0, 11.9 and 16.2 eV respectively, all of which are similar to that of hydrogen.  The neutral atoms of Mg$^\mathrm{o}$, Ca$^\mathrm{o}$ and Fe$^\mathrm{o}$ have ionization potentials of  7.6, 6.1 and 7.9 eV respectively so Mg$^+$, Ca$^+$ and Fe$^+$ can be produced by near-UV photons to which the HI zone is transparent.

Just as doing reverberation mapping for Fe~II is difficult because of the large size of the Fe~II emitting region, so Mg II reverberation mapping presents similar difficulties.  For example, \citet{Koratkar+Gaskell91} infer large sizes for the Mg II emitting regions, but with large uncertainties.  Because Mg~II showed a low amplitude of variability in the first two AGNs monitored in the {\it International AGN Watch} \citep{Clavel+91,Reichert+94}, it was subsequently decided not take UV spectra covering Mg~II in later campaigns since doing so would have doubled the amount of satellite observing time needed.

\subsection{Relationship to the surrounding dust}

\citet{Suganuma+06} find lags of hot dust emission in AGNs to be $\sim 3.5$ times the lag of H$\upbeta$ on average (see their Figure 32a).  Thus, the region with O$^\mathrm{o}$, Mg$^+$, Ca$^+$ and Fe$^+$ lies just inside the hot dust, and in the very outermost part of the BLR.

\section{The microphysics of Fe~II emission}

\subsection{Physical processes}

The basics of Fe~II line formation were derived by \citet{Netzer+Wills83} and \citet{Wills+85}.  Collisional excitation is responsible for the bulk of Fe~II emission but this alone gives poor agreement with observed spectra.  Many lines arising from high levels are required.  \citet{Netzer+Wills83} recognized that there were many wavelength coincidences between lines and that pumping and excitation of energy levels above 7 eV (self-fluorescence) improved agreement between calculated and observed Fe~II spectra. \citet{Johansson+Jordan84} recognized that Lyman $\upalpha$ fluorescence was another source of Fe~II emission and could explain unexpected UV lines seen in high-resolution UV spectra of RR Telescopii. \citet{Penston87} proposed that the same thing was going on in AGNs. Lyman $\upalpha$ fluorescence also produces strong Fe~II lines in the near-IR.  There is now good agreement between the predicted and observed strengths of these lines \citep{Garcia-Rissmann+12}.

Fluorescence requires overlapping lines.  The more energy levels and transitions that are included, the greater the overlap and the better the accuracy of calculations.  \citet{Wills+85} included 3407 transitions.  Recent models consider hundreds of thousands of transitions (see \citealt{Sakar+21} and references therein).  Providing atomic data for all these transitions has been an ongoing major challenge for atomic physicists.

\subsection{Turbulence and reddening}

The amount of overlap depends not only on the density of lines in the spectrum, but also on the amount of broadening.  The Doppler broadening is referred to as ``turbulence''.  \citet{Netzer+Wills83} recognized that it was easier to get fluorescence if the turbulent velocity of the gas was greater than the thermal velocities. More overlap between lines can then be achieved at lower optical depths.  \citet{Netzer+Wills83} noted that this particularly enhanced the ratio of UV to optical Fe~II emission. Recent calculations by \citet{Sakar+21} find that the overall shape and strength of the so-called ``Fe~II'' bump in the UV (also called the ``small blue bump'') can be reproduced with a turbulent velocity of $\thickapprox 100$ km s$^{-1}$. 

A big advantage of the GKN model of the BLR is that it readily provides high velocity dispersions as proposed by \citet{Baldwin+04}.  A high velocity dispersion {\it within} a single, traditional, optically-thick cloud is problematic, but high velocity dispersions {\it among} clouds spread out over light days, as in the GKN model, are natural.

An additional thing \citet{Wills+85} recognized was that it was much easier to explain the UV/optical Fe~II ratio if there was reddening.  They fit observations of Fe~II in AGNs with $E(B-V)$ of the order of 0.20.  As we will discuss in Section 13, reddening of this size is consistent with other indication.

\section{Fe~II Templates}

Good templates of Fe II emission are needed whether one is studying Fe II emission or trying to remove it from spectra to study other things. \citet{Sargent68} noted the similarity of line strengths in I~Zw~1 to those of 3C~273 and  \citet{Phillips77} showed that by empirically broadening the Fe~II spectrum of I~Zw~1 one could generally match Fe~II emission in AGNs with broader lines. \citet{Boroson+Green92} constructed a widely-used empirical Fe~II template by taking the optical spectrum of I~Zw~1 and removing all non-Fe~II lines.  For other AGNs it was assumed that the intensity ratios of the Fe~II lines were the same and only the widths changed.  \citet{Corbin+Boroson96} similarly created an empirical template for near-UV Fe~II emission. \citet{Vestergaard+Wilkes01} created an empirical template from {\it Hubble Space Telescope} spectra of I~Zw~1 covering the wavelength range $\uplambda\uplambda$1250 - 3090.  They give an excellent discussion of the issues involved in creating Fe~II templates in general (see also \citealt{Park+21}).  Since then additional empirical templates have been created.  For example, \citet{Veron-Cetty+04} identified a large number of Fe~II lines in I~Zw~1.  

In parallel with these empirical approaches, progress has been made in the theoretical modelling of Fe~II (see previous section).  While we are not at the stage yet of being able to produce a purely theoretical Fe~II spectrum (lack of sufficient high quality atomic data alone is a limitation), theoretical modelling has allowed the construction of ``semi-empirical'' templates that combine theoretical predictions with observations.

\citet{Tsuzuki+06} used a combination of {\it Cloudy} models and observations of I~Zw~1 to created a combined UV and optical template.  An important thing \citet{Tsuzuki+06} do is to recognize the need to include internal reddening and to deredden spectra.  For I~Zw~1 they get $E(B-V) = 0.09$ from the ratio of O~I $\uplambda$1304/$\uplambda$8446 when they assume an SMC-like reddening curve.  However, the reddening curve of the majority of AGNs is much flatter than an SMC curve (\citealt{Gaskell+04}; \citealt{Czerny+04}; \citealt{Gaskell+Benker07}).  Using the average AGN reddening curve of \citet{Gaskell+Benker07} gives $E(B-V) = 0.23$, which similar to the reddening of a typical AGN as shown in Figure 11 (see next section).  The semi-empirical optical template of \citet{Kovacevic+10} gives good fits to optical spectra (as illustrated, for example, by the residuals in Figure 9) by allowing groups of multiplets to vary independently.  As we have shown in Figure 11, much of the variation in the ratios of the different groups is due to reddening. 
     
\citet{Garcia-Rissmann+12} have constructed a semi-empirical Fe II template that included near-IR Fe~II emission.  Many lines are explained well by a theoretical model, but other show substantial differences and so were empirically corrected to match the I~Zw~1 spectrum.  The empirical template so derived gives a good fit to the spectrum of the NLS1 Akn 564.

The majority of templates have been based on I~Zw~1. Recently, \citet{Park+21} have produced a template of the $\uplambda$4000 -- $\uplambda$5600 wavelength range using {\it HST} observations of Mrk~493 instead. Mrk~493 has the advantage of having narrower Fe~II lines than I~Zw~1.  \citet{Park+21} also compare and discuss other Fe~II templates and the effects of using different templates.

In summary, good progress has been made in making templates from the UV to the near-IR.  Because of the limitations of knowledge of the atomic physics, templates are going to continue to have to be semi-empirical for the foreseeable future.

\section{Reddening}

It has been very common to assume that internal reddening in AGNs is negligible and to only consider Galactic reddening due to dust in the solar neighborhood.  There has, however, long been evidence that internal reddening is far greater than Galactic reddening (see \citealt{Gaskell17} for a review).  When using an Fe~II template to fit Fe~II lines, the spectrum first needs to be corrected for reddening. In their Fe~II template \citet{Kovacevic+10} separate out the main optical multiplets by their lower terms: $^4$F, $^4$S, and $^4$G (see their Figure 1a).  They present fits of these groups to a large number of SDSS AGNs.  The ratio of the strengths of these groups varies from AGN to AGN.  \citet{Kovacevic+10} note correlations with FWHM H$\upbeta$ and the H$\upalpha$/H$\upbeta$ ratio.  These could indicate that systematic differences in the ratios of the strengths of the different multiplet groups are a consequence of different physical differences in the AGNs but another possibility  \citet{Kovacevic+10} mention is that the differences are due to reddening, since the groups differ in mean wavelength.  We have explored this possibility quantitatively.

\begin{figure}
\centering 
\includegraphics[height=0.47\textwidth,angle=0]{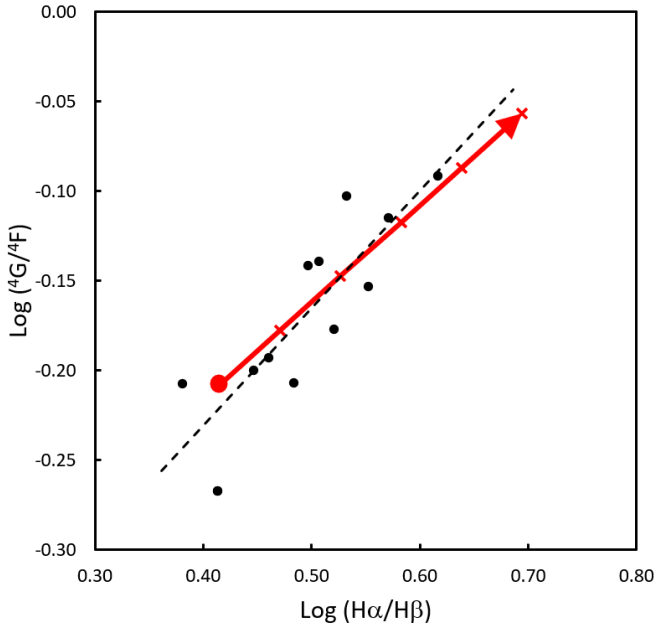}
\caption{Average ratios of the Fe~II $^4$G and $^4$F multiplets as fit by \citet{Kovacevic+10} as a function of the total line flux Balmer decrement for 127 SDSS AGNs.    The red arrow shows the reddening vector with tick marks every 0.1 in $E(B-V)$.  The unreddened H$\upalpha$/H$\upbeta$ ratio has been taken to be 2.6 (see \citealt{Gaskell17}) and the unreddened $^4$G/$^4$F ratio is from the fit to the points. The reddening vector is from the mean AGN reddening curve of \citet{Gaskell+Benker07}.  The dotted black line is a linear regression of log $^4$G/$^4$F on log H$\upalpha$/H$\upbeta$.}
\end{figure}

The $^4$G and $^4$F multiplet groups are the main contributors to the Fe~II $\uplambda$4570 and $\uplambda$5300 blends respectively. In Figure 11 we show the correlation between the mean intensity ratio of the $^4$G and $^4$F groups as a function of the Balmer decrement for the 127 SDSS AGNs with a low enough redshift for H$\upalpha$ to be measured.  We have binned the AGNs into equal-sized bins by H$\upalpha$/H$\upbeta$.  It can be seen in the figure that there is a strong clear correlation of the ratio of intensities of the $^4$G and $^4$F groups with H$\upalpha$/H$\upbeta$.  Furthermore, the slope of the trend is in excellent agreement with that expected for intrinsic reddening.  The reddening vector we show in Figure 11 is from the mean AGN reddening curve of \citet{Gaskell+Benker07}, but over this wavelength range, all reddening curves are similar.  The mean reddening of these 127 AGNs is $E(B-V) \sim 0.2$.  SDSS AGNs are colour selected and biased towards blue AGNs.  Reddening of AGNs chosen in other ways will be greater and the application of reddening correction before fitting Fe~II template is all the more important.

\section{Eddington ratio: the driver of eigenvector 1}

\begin{figure}
\centering 
\includegraphics[width=0.48\textwidth,angle=0]{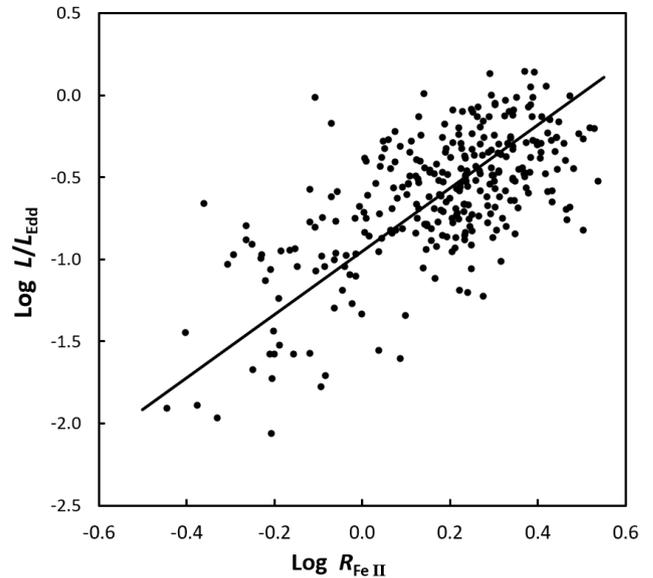}
\caption{The Eddington ratio as a function of $R_{\mathrm{FeII}}$.  The line is a least-squares bisector fit.}
\end{figure}

\citet{Boroson+Green92} found that the main components of EV1 are inverse correlations of $R_{\mathrm {FeII}}$ with the FWHM of H$\upbeta$ and the strength of the [O III] lines. They speculated that the physical driver of EV 1 could be the Eddington ratio, $L/L_{\mathrm{Edd}}$.   \citet{Wandel+Boller98} showed how a high $L/L_{\mathrm{Edd}}$ would indeed produce narrow BLR lines. Since $R_{\mathrm {FeII}}$ is part of EV1, we would therefore expect it to show a correlation with $L/L_{\mathrm{Edd}}$. Black hole masses can be readily estimated by the single-epoch spectrum method of \citet{Dibai77}, where the mass is found from $R v^2$, with the effective velocity, $v$, being taken to be the FWHM, and the effective radius, $R$, being estimated from $L^{1/2}$.  The Eddington ratio is therefore proportional to $L^{1/2} v^{-2}$. If EV1 is driven by $L/L_{\mathrm{Edd}}$, the components of EV1 should be better correlated with $L/L_{\mathrm{Edd}}$ than with each other. We can test this using the SDSS sample of 300 AGNs measured by \citet{Kovacevic+10}.  There is an uncertain scale factor in calculating the Eddington ratio.  This is because of the uncertainty in the bolometric correction and in determining the mass from the spectra of the BLR, but we have taken $\log L/L_{\mathrm{Edd}}$ to be $2.5 \log L_{5100} - 2 \log \mathrm{FWHM} -16.3$, where $L_{5100}$ is the monochromatic luminosity at $\uplambda$5100 in ergs s$^{-1}$ Hz$^{-1}$. The uncertainty in the scale factor has no effect on our discussion.

The square of the  correlation coefficient between $R_{\mathrm{FeII}}$ and FWHM is $r^2 = 0.27$ while for $R_{\mathrm{FeII}}$ and $L/L_{\mathrm{Edd}}$ we get  $r^2 = 0.42$.  We show the latter correlation in Figure 12. Note that the sparsity of points to the lower left is an artifact of \citet{Kovacevic+10} restricting their sample to where they could reliably measure Fe II.  The equivalent width of [O III] also shows a somewhat stronger correlation with $L/L_{\mathrm{Edd}}$ than with FWHM, although neither correlation is as strong as the correlation with Fe~II.\footnote{The weaker correlation for [O III] is because [O III] is correlated with luminosity \citep{Steiner81}.  Also, \citet{Heard+Gaskell16} find that both the equivalent widths of the [O III] lines and the ratio of their intensity to broad H$\upbeta$ are correlated with the reddening of the BLR in a manner which is consistent extinction by dust interior to the NLR.}

\begin{figure}
\centering 
\includegraphics[width=0.49\textwidth,angle=0]{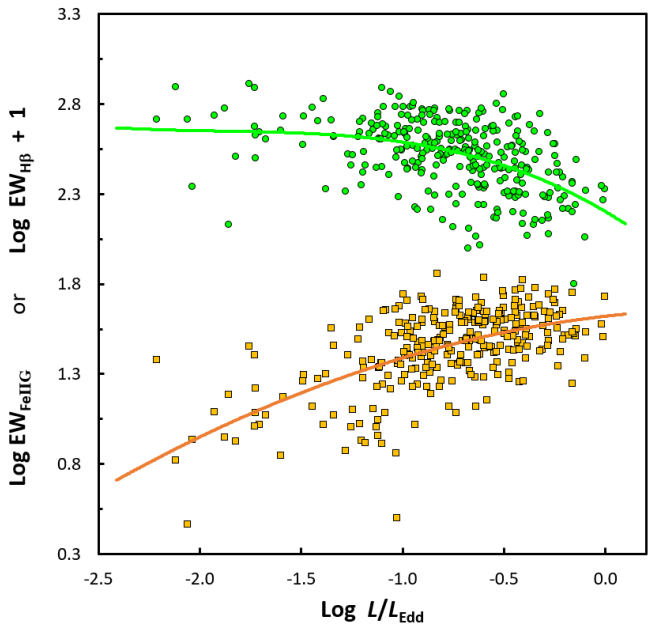}
\caption{The variation of the equivalent widths of the narrow core of broad H$\upbeta$ (green circles in the upper part of the diagram) and of the $^4$G blend of Fe~II at $\uplambda$5300 (brown squares in the lower part of the diagram) as fit by \citet{Kovacevic+10} as a function of Eddington ratio.  The H$\upbeta$ points have been offset for plotting convenience.  The curves for H$\upbeta$ and Fe~II are respectively third and second order polynomials.}
\end{figure}

From $R_{\mathrm{FeII}}$ one can predict $L/L_{\mathrm{Edd}}$ to 0.32 dex.  There are good reasons to believe that the intrinsic scatter in the underlying correlation must be less than this.  The biggest reason is that the optical luminosity of AGNs is variable.  The Eddington ratio can thus vary on timescales from days to weeks.   After subtraction of the contamination of host-galaxy starlight, optical variability of a factor of ten is not uncommon in AGNs with low average Eddington ratios.  A second reason why the intrinsic scatter must be less than in Figure 11 is that the mass determined by the Dibai method is uncertain by $\sim \pm 0.28$ dex (see, for example, \citealt{Bochkarev+Gaskell09}).  Further scatter is introduced by neglecting reddening, which will introduce scatter into $L$, and by variation in metallicity, which can cause variation in the strength of the Fe~II emission, especially when Fe~II is not too strong.

Given the good correlation with $L/L_{\mathrm{Edd}}$ despite the expected sources of scatter, $R_{\mathrm{FeII}}$ seems to be a useful proxy for studying the Eddington ratio of AGNs. It has the advantage of not being dependent on $L_{5100}$ and thus not being influenced by the inevitable biases in $L$ that are present in any magnitude-limited or signal-to-noise-ratio-limited sample.

\section{The cause of weak Fe~II}

\citet{Gaskell+81} argued that the strengths of the broad lines of the  refractory elements silicon, aluminium, magnesium and iron in AGNs ruled out the order-of-magnitude depletions that would be expected if grains were present in the BLR.  Thus little or no dust can be present in the BLR when these lines were strong.  However, as we have noted, Fe~II varies considerably in strength and so depletion is possible.  \citet{Osterbrock84} proposed that this was indeed what was happening and that depletion of Fe onto dust grains was the cause of the weak Fe~II in broad-line radio galaxies.

\section{Metallicity}

There is a mass-metallicity relationship for galaxies as a whole and the metallicity increases for stars towards the centre of a galaxy (see, for example, \citealt{Kobayashi+Arimoto99}). The metallicity of the stars goes up to [Fe/H] $\sim +0.3$ or  +0.5 in the centres of the most massive galaxies (i.e., up to two or three times solar). What matters for feeding AGNs is the gas-phase abundance and this might be another +0.2 dex higher than stellar abundances.  Metallicity variations of factors of several in galactic nuclei are to be expected but extremely high metallicities (20 to 80 times solar, say) are ruled out.  We have already noted that models with solar abundances can reproduce a typical $R_{\mathrm{FeII}}$, so, extremely high abundances of iron are not required.

We should note that the strength of Fe~II emission will {\it not} be linear with the iron abundance when the Fe~II lines are strong.  The great strength of Fe~II lines in AGNs means that {\it Fe~II is the main coolant} of the gas in which they are produced. When lines of an element are the dominant coolant, their strength is relatively insensitive to the abundance. This is because thermal equilibrium requires that the energy removed by the collisionally-excited lines balances the input energy.  If the abundance of the element whose lines are dominating the cooling is low, the gas compensates by raising the temperature to increase the emission.  The equivalent width of Fe~II will thus be insensitive to the abundance when Fe~II is the main coolant.  It can be seen in the lower half of Figure 13 that the equivalent width of the $^4$G $\uplambda$5300 Fe~II blend does appear to saturate at high Eddington ratios.\footnote{The $^4$F and $^4$S blends of \citet{Kovacevic+10} show similar correlations but with more scatter since they are blended with He II $\uplambda$4686 and [O III] plus the red wing of H$\upbeta$ respectively. $^4$F is also the blend least affected by reddening -- see section 13.}

\section{The weakening of H$\upbeta$}

The correlation of $R_{\mathrm{FeII}}$ with the FWHM of H$\upbeta$ (see Figure 2) is a key part of EV1. However, it is a ratio and when a ratio increases, this can be due to the numerator increasing or the denominator decreasing. \citet{Gaskell85} found that the equivalent width of H$\upbeta$ declines with FWHM. \citet{Kovacevic+10} separated broad H$\upbeta$ into two components: a narrow core which has a similar width to Fe~II, and a broader component. An example it shown in the upper half of Figure 9. The narrow core will correspond to the gas emitting Fe~II in the outer BLR.  The broader component will be correspond to the higher ionization inner regions of the BLR. The narrow core shows a correlation  with the Eddington ratio ($r^2 = 0.23$) as shown in the upper half of Figure 13; the broad component (not plotted) does not ($r^2 = 0.004$).  

We believe that the weakening of H$\upbeta$ with increasing $L/L_{\mathrm{Edd}}$ is a consequence of the different ways in which the Fe~II lines and H$\upbeta$ are excited.  Fe~II is produced mostly by collisions. As we have noted, it is the major coolant in the Fe$^+$ zone and its strength depends on the rate of {\it heating}.  H$\upbeta$, on the other hand, is produced mostly by recombinations and so its strength depends on the rate of {\it ionizations}. The ratio of heating to ionization is different for lower energy ionizing photons with energies, $E$, around the 13.6 eV ionization threshold for hydrogen than for higher energy photons ($E > 0.1$ keV, say).  Lower-energy ionizing photons ionize but do not do as much heating compared with higher-energy photons. The heating-to-ionization ratio is higher for higher-energy photons because a single photon has more energy and this energy is dissipated through Auger electrons, etc.   

Figure 11.5 of \citet{Osterbrock+Ferland06} shows the total cross section for gas as a function of photon energy for all processes. From this figure we can see that for a metallicity of two or three times solar (the most likely case in an AGN), photoelectric absorption by heavy elements exceeds that by hydrogen for photon energies greater than 100 eV because the photoionization cross section of hydrogren falls off with frequency as $\nu^{-3.5}$.  Only high-energy photons will survive to make it far into the gas with Fe$^{+}$.  The gas is heated and hydrogen is kept partially ionized by high-energy photons. 

There are two ways to increase the relative rate of absorption of photons by heavy elements relative to absorption by neutral hydrogen. The first is to have more soft X-ray photons; the second is to have a higher abundance of heavy elements.  Doing either or both of these things will raise the ratio of heating to ionization and hence the raise the ratio of Fe~II/H$\upbeta$.  Since the intensity of Fe~II is fixed when it is the major coolant, the intensity of H$\beta$ will go down.  If we consider just the metal abundance, an alternative way of looking at this is to recognize that if we increase the metallicity, the {\it hydrogen} column density in which photons are stopped by heavy elements decreases. The Fe~II strength will stay the same, because it has to carry away the energy, but because the column density of hydrogen is less, there will be less hydrogen to be collisionally ionized.

\section{AGN downsizing: the cause of the correlations with radio properties and galaxy type}

As noted earlier, correlations of optical Fe~II emission with radio properties were discovered early on in the 1970s.  Naturally, these led to consideration that there might be a physical link between Fe~II emission and radio emission. This motivated some models for Fe~II emission.  We propose instead that {\it the correlation of optical Fe~II strength is a consequence of black hole ``downsizing''.}

It has long been known \citep{Schmidt68} that AGN activity was vastly greater at high redshifts with a peak around $z \sim 2$.  Not only AGN activity but also star formation \citep{Madau+96} peaked around this so-called ``cosmic high noon'' at $z \sim 2$.  The decline in AGN activity since then is referred to as ``AGN cosmic downsizing''.  The most massive galaxies that were vigorously star forming at $z \sim 2 - 3$ are now mostly ``red and dead''.  Because of the general lack of gas, their very massive black holes are now accreting at much lower Eddington ratios than they were at $z \sim 2 - 3$.  On the other hand, star formation continues today in lower mass galaxies along with the feeding of their correspondingly lower mass black holes.

\citet{Franceschini+98} made the important discovery that the observed radio luminosity at 5 GHz, $L_{5GHz}$, was strongly correlated with the mass, $M_{bh}$, of the black hole and with $L_{5GHz} \propto M_{bh}^{2.5}$.  There are various biases and selection effects which weaken the fundamental correlation, especially for flat-spectrum radio sources that are preferentially seen with the jet close to our line of sight (see, for example discussion in \citealt{Jarvis+McLure02} and \citealt{Liu+06}), but \citet{Liu+06} conclude that ``mass plays a dominant role in producing the jet power, compared with the Eddington ratio.''  A large   optically-selected sample \citep{McLure+Jarvis04} gives a consistent result: radio-loud AGNs have more massive black holes than radio-quiet AGNs.  This is consistent with the old result \citep{Setti+Woltjer77} that radio-loud AGNs tend to be in more massive galaxies (i.e., ellipticals) than radio-quiet AGNs which are most commonly found in spirals. When considering the reported correlations between the strength of Fe~II and the observed radio properties, the biases can mostly be ignored because what matters is what is selected by the optical observers based on {\it observed} radio properties.\footnote{For example, in the past, optical observers did not choose targets based on fluxes corrected for Doppler boosting.} Also, biases are least for steep-spectrum radio sources, such as the radio galaxies considered by \citet{Grandi+Osterbrock78} or the extended radio structure AGNs considered by \citet{Miley+Miller79}. Figure 6 of \citet{Franceschini+98} tells us that {\it if one chooses an AGN with a high observed power at 5 GHz, one is selecting a very high mass black hole.}

Radio jets can be produced at very low Eddington ratios (so-called non-thermal AGNs -- see \citealt{Antonucci12}) because the jets are believed to be powered by the Blandford-Znajek process through the extraction of spin energy from the black hole \citep{Blandford+Znajek77}. A sample of AGNs selected on radio power, especially on low-frequency power and hence on the power of the extended emission in the lobes, will therefore include objects with low Eddington ratios.  As we have seen, low Eddington ratios produce lower Fe~II emission.  Powerful steep-spectrum radio-loud AGNs will thus tend to have weaker Fe~II emission as observed.  

The radio emission in flat-spectrum sources is greatly enhanced by Doppler boosting. This brings intrinsically weaker sources into flux-limited samples and gives a strong bias towards face-on AGNs.  A sample of flat-spectrum sources that is flux limited in the radio will have much lower intrinsic jet power than a steep-spectrum sample (see \citealt{Liu+06}).  It will {not} be biased towards high black hole masses the way a steep-spectrum sample is. A sample of flat-spectrum, radio-loud AGNs thus resembles a radio-quiet sample.  It will have lower black hole masses, higher Eddington ratios, and hence stronger Fe~II emission.

AGN downsizing predicts that $R_\mathrm{FeII}$ will increase with increasing redshift.  \citet{Kovacevic+10} find this to be the case (see their Figure 18), but their magnitude-limited sample only includes high luminosity sources at higher redshifts and selecting higher-luminosity sources biases the sample to higher Eddington ratios.  Careful allowance for selection biases and larger, deeper samples are needed for evaluating how the Eddington ratio changes with redshift.

\section{Some Future directions}

If the overall picture we have presented here is correct, a coherent picture of Fe~II emission in AGNs and explanations for various aspects of ``the Fe~II problem'' is starting to form.  There is much that needs testing however.  We conclude with mentioning some possible future directions for research.

\subsection{Reverberation mapping}

Fe~II reverberation has been a struggle.  Although reverberation mapping campaigns studying the Balmer lines have been carried out for nearly half a century, it has only been within the last decade that there has been some success with Fe~II reverberation.  Our analysis of even the best reverberation-mapping campaigns to date shows that much improvement is needed.  The main shortcoming can now be seen to be the short duration of the campaigns compared with the Fe~II lags.  We recommend that future campaigns be at least ten times longer than the expected Fe~II lag to reduce the biases identified discussed in section 9.4.  The best way to do this is to choose high declination AGNs so that there are no annual gaps.  Longer campaigns also improve the chances of catching strong continuum variability.

\subsection{Modelling Fe~II emission from realistic BLRs}

In addition to ongoing work improving modelling of Fe~II emission by including more atomic details and better atomic data, work is needed incorporating the improved Fe~II modelling into more realistic models of the BLR that satisfy constraints set by line profiles and velocity-resolved reverberation mapping.

\subsection{Causes of scatter in EV1}

Depletion of refractory elements onto grains is something that changes the gas-phase abundance of iron by orders of magnitude.  However, changes in the photoionization conditions, such as the gas density, column density, and spectral energy distribution make a difference too.  The cause of scatter in EV1 needs exploring through models and analysis of data sets.  Care is needed in considering selection effects.

\subsection{Dust and Fe~II}

The destruction of dust in AGN environments is an important detail that needs more investigation, as does the role of reddening.

\subsection{Downsizing}

The ``downsizing'' explanation of the reported relationships between Fe~II emission and radio properties needs to be tested with well-defined samples and allowance for selection effects.  A consequence of AGN downsizing, which is quite pronounced for $z  < 1$ is that Fe~II emission becomes stronger on average with increasing redshift.  This is indeed what is found by \citep{Kovacevic+10}. However,  because higher Eddington-ratio sources are brighter for a given black hole mass, careful allowance needs to be made for selection effects. As we have shown, $R_{\mathrm{FeII}}$ is a powerful estimator of $L/L_{\mathrm{Edd}}$ that has the advantage of not using $L$ and $M$ directly.  It should therefore be a powerful tool for studying the evolution of $L/L_{\mathrm{Edd}}$ over time.

\subsection{An ``iron chronometer''?}

Iron is produced in supernovae.  For Fe produced in type Ia supernovae one has to wait until there has been enough time to form a white dwarf.   There could therefore be a delay of $\sim 1$ Gyr between the production of $\upalpha$ elements, such as Mg, in core-collapse supernovae and Fe enrichment.   At $z = 5.5$ the universe is only one billion years old, so the abundance ratio of Fe/Mg is potentially a chronometer for star formation history.  Unfortunately, measuring the abundance of Fe in AGN spectra is difficult.  A lot needs to be well understood before we can have a reliable chronometer.  This is well illustrated by the results of \citet{Verner+Peterson04}.  They find that the fraction of AGNs with high ratios of Fe~II/Mg~II {\it increases} noticeably at $z > 2$, which is the opposite of what is expected from supernova enrichment if the Fe~II/Mg~II ratio is measuring the abundance of Fe.  They also note that the increase for $z > 2$ is for the most luminous AGNs. As \citet{Verner+09} comment,  this luminosity dependence contradicts the na\"{i}ve idea that Fe~II/Mg~II is an abundance indicator.

We suggest that the decrease in Fe~II/Mg~II for the most luminous AGNs for $z < 2$ is an effect of AGN downsizing and the correlation of the strength of Fe~II with Eddington ratio.  $z \sim 2$ is the redshift at which the highest luminosity AGNs begin to turn off (see Figure 4 of \citealt{Ikeda+11}).  This illustrates the need for more work the factors governing Fe~II strength.

\section{Conclusions}

Our main conclusions are:

(1) Despite earlier doubts it is now clear that Fe~II emission in AGNs is produced by photoionization.

(2) The self-shielding BLR model of \citet{Gaskell+07} predicts that Fe~II is predominantly emitted at a radius twice that of H$\upbeta$.  This is similar to the radius O~I, Mg~II and Ca~II are emitted at.

(3) The relative line widths of Fe~II and H$\upbeta$ support the hypothesis that Fe~II is predominantly emitted at twice the radius of H$\upbeta$.

(4) The main difficulties in doing Fe~II reverberation mapping are (a) that the large size of the Fe~II emitting region causes the light travel time to wash out variability of the Fe~II lines, (b) that high Eddington ratio objects have continuum variability that is of low amplitude and rapid compared with lower Eddington ratio AGNs. and (c) that anomalies in responses in reverberation mapping are on a similar timescale to Fe~II lags.

(5) After allowing for the ``washing out'' of variability because of the long delays, Fe~II responds just as strongly to continuum variations as H$\upbeta$.

(6) Reverberation-mapping campaigns that are short compared with the expected Fe~II lags and suffer from low signal-to-noise ratios will systematically underestimate sizes, as shown by \citet{Welsh99}.

(7) The best quality reverberation mapping to date shows Fe~II coming from twice the radius of H$\upbeta$ in agreement with the predictions of the BLR model of \cite{Gaskell+07} and Fe~II line widths.

(8) The region emitting Fe~II, Mg~II, Ca~II and O~I extends out to the warm dust, which is  3.5 times further out on average than H$\upbeta$.

(9) The relative velocities of clouds in the \citet{Gaskell+07} model produce the Doppler broadening needed to enhance the production of Fe~II.

(10) The correlation of the ratio of strengths of Fe~II multiplets at different wavelengths with the Balmer decrement implies reddenings of the order of $E(B-V) \sim 0.2$ for SDSS colour-selected AGNs.  This reddening increases the ratio of UV/optical Fe~II emission and makes explaining Fe~II emission easier.

(11) The ratio of intensity of optical Fe~II emission to the intensity of H$\upbeta$,  $R_{\mathrm {FeII}}$, is a good proxy for the Eddington ratio.

(12) The locking up of Fe in grains in the outer BLR is a likely cause of weak Fe~II emission.

(13) Because Fe~II is a major coolant, the strength of Fe~II is insensitive to [Fe/H] when Fe~II lines are strong.  

(14) The reported correlations of Fe~II strength with radio type and host galaxy type are a consequence of higher mass black holes having lower Eddington ratios in the present day universe while lower mass black holes still have high Eddington ratios (i.e., ``AGN cosmic downsizing'').

(15) There are many problems to be overcome in using Fe~II emission in AGNs to study of changes in [Fe/H] with redshift.

\section*{Acknowledgments}

NT, BT and AS carried out their work under the auspices of the Science Internship Program (SIP) of the University of California at Santa Cruz.  We wish to express our appreciation to Raja GuhaThakurta for his excellent leadership of the SIP program. CMG wishes to express his appreciation to the organizers of the 13th Serbian Conference on Spectral Line Shapes for the opportunity to present this work.  We are grateful to Jelena Kova{\v{c}}evi{\'c} Doj{\v{c}}inovi{\'c} for helpful comments.

\bibliography{Martin_Gaskell}%

\end{document}